\documentclass[%
 aip,
 amsmath,amssymb,
 reprint,%
]{revtex4-1}

\usepackage{graphicx}
\usepackage{dcolumn}
\usepackage{bm}

\usepackage[utf8]{inputenc}
\usepackage[T1]{fontenc}
\usepackage{mathptmx}
\usepackage{etoolbox}
\usepackage{todonotes}

\makeatletter
\def\@email#1#2{%
 \endgroup
 \patchcmd{\titleblock@produce}
  {\frontmatter@RRAPformat}
  {\frontmatter@RRAPformat{\produce@RRAP{*#1\href{mailto:#2}{#2}}}\frontmatter@RRAPformat}
  {}{}
}%

\usepackage{xr}
\usepackage{natbib,hyperref}
\makeatletter
\newcommand*{\addFileDependency}[1]{
  \typeout{(#1)}
  \@addtofilelist{#1}
  \IfFileExists{#1}{}{\typeout{No file #1.}}
}
\makeatother
\newcommand*{\myexternaldocument}[1]{
    \externaldocument{#1}
    \addFileDependency{#1.tex}
    \addFileDependency{#1.aux}
}
\myexternaldocument{si}

\begin{document}

\preprint{AIP/123-QED}

\title[Invasion and Interaction Determine Population Composition in an Open Evolving System]{Invasion and Interaction Determine Population Composition in an Open Evolving System}
\author{Youngjai Park}
\affiliation{Department of Physics, Inha University, Incheon 22212, Korea}
\affiliation{Asia Pacific Center for Theoretical Physics (APCTP), Pohang 37673, Korea}
\affiliation{Department of Applied Physics, Hanyang University, Ansan 15588, Korea}

\author{Takashi Shimada}
\affiliation{Mathematics and Informatics Center, The University of Tokyo, Tokyo 113-8656, Japan}
\affiliation{Department of Systems Innovation, Graduate School of Engineering, The University of Tokyo, Tokyo 113-8656, Japan}
  
\author{Seung-Woo Son}
\email{sonswoo@hanyang.ac.kr}
\affiliation{Asia Pacific Center for Theoretical Physics (APCTP), Pohang 37673, Korea}
\affiliation{Department of Applied Physics, Hanyang University, Ansan 15588, Korea}
\affiliation{Center for Bionano Intelligence Education and Research, Hanyang University, Ansan 15588, Korea}

\author{Hye Jin Park}
\email{hyejin.park@inha.ac.kr}
\affiliation{Department of Physics, Inha University, Incheon 22212, Korea}
\affiliation{Asia Pacific Center for Theoretical Physics (APCTP), Pohang 37673, Korea}

\date{\today}

\begin{abstract}
It is well-known that interactions between species determine the population composition in an ecosystem.
Conventional studies have focused on fixed population structures to reveal how interactions shape population compositions.
However, interaction structures are not fixed, but change over time due to invasions.
Thus, invasion and interaction play an important role in shaping communities.
Despite its importance, however, the interplay between invasion and interaction has not been well explored.
Here, we investigate how invasion affects the population composition with interactions in open evolving systems considering generalized Lotka-Volterra-type dynamics. 
Our results show that the system has two distinct regimes. 
One is characterized by low diversity with abrupt changes of dominant species in time, appearing when the interaction between species is strong and invasion slowly occurs. 
On the other hand, frequent invasions can induce higher diversity with slow changes in abundances despite strong interactions.
It is because invasion happens before the system reaches its equilibrium, which drags the system from its equilibrium all the time.
All species have similar abundances in this regime, which implies that fast invasion induces regime shift.
Therefore, whether invasion or interaction dominates determines the population composition.
\end{abstract}

\maketitle

\begin{quotation}
An ecosystem consists of many interacting species.
The generalized Lotka-Volterra equation has described how such interactions determine the abundances of species.
Most conventional studies, however, have focused on fixed interaction structures, while real-world ecosystems evolve by constantly introducing new species.
Once new species invade a system, the interaction structure changes because the new species induce new interactions.
In this paper, we capture the dynamics of the interaction structure considering an open evolving network. Species and pairwise interactions are represented by nodes and links in the network. 
The interplay between invasion and interaction will shape the abundances of species.
If the invasion occurs frequently, new species will invade a system before the system reaches its equilibrium. 
As the interaction strength governs the equilibration time, the interaction strength as well as the invasion rate plays an important role in the dynamics.
Examining the abundance distributions of species for various invasion rates and interaction strengths, we find that reducing invasion rate and increasing interaction strength affect in a similar way on the abundance distributions.
Furthermore, we figure out the role of the invasion rate and the interaction strength by measuring the correlation between abundances and species properties such as age and the incoming degree strength in the given interaction structure.
It sheds light on that invasion as well as interaction plays an important role to determine the population composition -- diversity and abundance distribution -- in an open evolving system. 
\end{quotation}

\section{Introduction} \label{sec:intro}
In an ecosystem, many species interact with each other.
Some species compete with others for shared resources, inducing adaptation or death from the competition. 
On the other hand, flowers provide food for insects, with the help of reproduction from them. 
As interactions between species affect their death and reproduction, population compositions are determined from them.
To understand how such interactions affect populations, the generalized Lotka-Volterra equation has been studied~\cite{lotka1920analytical,volterra1928variations}. 
The equation describes the abundance dynamics of species taking interactions into account and has successfully explained the real world~\cite{xue2017coevolution,farahpour2018trade,sidhom2020ecological}.
However, the literature has focused on the situation where the interaction structures are fixed, while invasions alter the structures all the time in nature.

An ecosystem is an open system that has biological invasions.
A newly invading species interacts with resident species resulting in changes in the interaction structure, and as a consequence, the population composition changes.
If there is no invasion at all, population dynamics with a non-changing interaction structure can be analyzed by investigating fixed points of the dynamical system~\cite{sahney2010links,allesina2012stability,goyal2018diversity,lin2019spatial,pettersson2020predicting,pettersson2020stability,bunin2017ecological,park2021stochastic}. 
The system evolves towards the stable fixed points as time goes.
However, the open evolving system with new invaders continues to change interaction structures over time. 
Here, another timescale which comes from the invasion process is involved in population dynamics, and thus the equilibration is not guaranteed, where the previously well-developed methodology could not be applied.

To implement the evolving interaction structure, we introduce an evolving network, where nodes represent species and links are pairwise interactions~\cite{taylor1988consistent,taylor1988construction,shimada2002self,tokita2003emergence,mathiesen2011ecosystems,shimada2014universal}. 
In the ecosystem, the interactions are directional with weights and signs. 
The weight indicates how strong the interaction is and the sign does the type of the interaction such as competition, facilitation, mutualism, parasitism, and so on~\cite{pringle2016orienting,ogushi2017enhanced,schob2018evolution,losapio2021network,park2021sign}. 
Furthermore, the abundance of each species is denoted as a property of the node.
Interactions finally change the abundance of all species and even can cause extinction of certain species.
In this case, extinct species nodes are removed from the system with the connecting/connected links together.
Once a new species comes (invades) into the system, a new species node and connecting links (new interactions) are added.

Considering the evolving interaction networks with generalized Lotka-Volterra-type dynamics, we investigate how both invasion and interaction determine the population composition. 
As the system has two timescales involved in the invasion and equilibration processes, we manipulate the invasion rate and interaction strength.
Varying these two parameters, we perform simulations of population dynamics and find two distinct patterns in the observed abundance, determined by the competition between invasion and interaction timescales.

Fast invasion helps the system keep high diversity pushing the settled resident species. 
It suppresses the emergence of dominant species that take over almost the population if there is enough time to evolve. 
The high-diversity populations appear if the invasion is fast, even for the quite strong interaction condition, which makes equilibration fast.
However, when interactions are strong enough to overcome the invasion effect, the dominant species finally appear.
The decreasing invasion rate and increasing interaction strength are effectively the same in the sense of diversity. It only changes the time-scale factor.
Furthermore, the abundance distributions of each species are similar, which implies that the dominant factor between the invasion rate and the strength of the interaction determines the population composition.

This paper is organized as follows: The Lotka-Volterra type equation with resource limitation is portrayed in Sec.~\ref{subsec:dynamics}. 
In Sec.~\ref{subsec:network}, the scheme of an open evolving network system is described.
Section~\ref{sec:result} is focused on the correlation between abundance and a network property to infer the boundaries among different regimes that show different population compositions. 
A further discussion is in Sec.~\ref{sec:discussion}.

\section{Model} \label{sec:model}
Interactions between species determine the abundance of species telling who will dominate and who will go extinct. 
On the other hand, invasions perturb the existing population composition because new interactions reshape the interaction structure.
To investigate both invasion and interaction effects on population composition, we construct an open evolving model with generalized Lotka-Volterra-type dynamics.

\subsection{Abundance dynamics} \label{subsec:dynamics}
    \begin{equation}
    \frac{dx_i}{dt} = r_i \; x_i \; \left( 1 - \frac{\sum_j^S {w_{ij} \; x_j}}{K_i} \right) \: ,
    \label{eq:compLV}
    \end{equation}
where $x_i$ is the abundance of species $i$. 
The intrinsic growth rate and the carrying capacity of species $i$ are denoted by $r_i$ and $K_i$, respectively.
The effect of the interaction from species $j$ to species $i$ is $w_{ij}$.
When species $j$ affects the abundance of species $i$, $w_{ij} \neq 0$. 
Otherwise, the interaction weight $w_{ij}$ is zero.
If there are no interactions at all, Eq.~\eqref{eq:compLV} is equivalent to the logistic growth equation with $w_{ii}=1$. 

The generalized Lotka-Volterra equation has captured the behavior of ecosystems well~\cite{farahpour2018trade,sidhom2020ecological}.
When it comes to open evolving systems, however, sometimes populations grow without a bound which is unrealistic when the amount of common resources such as space and shared food is finite so-called resource limitation. 
Thus, we modify the generalized Lotka-Volterra equation so that the system has the bounded total population size $K$ as follows
    \begin{equation}
    \frac{dx_i}{dt} = G_i(\textbf{x}) \; x_i \; \left( 1-\frac{\sum_j^{S(t)}{x_j}}{K} \right) \; + \; D_i(\textbf{x}) \; x_i \: ,
    \label{eq:dxdt}
    \end{equation}
with the abundance vector $\textbf{x} = \{ x_i \}$ in which elements consist of the abundances of all species. 
The first and the second terms on the right-hand side describe birth and death processes, respectively.
Thus, birth occurs only when the total population does not reach the total carrying capacity $K$.
The growth and death rates of species $i$, $G_i(\textbf{x})$ and $D_i(\textbf{x})$, change depending on the interaction weights and the abundance of the interacting species:
    \begin{equation}
    G_i(\textbf{x})=\sum_{j}{w_{ij}^+ \; x_j} \: ,
    \label{eq:g_x}
    \end{equation}
and
    \begin{equation}
    D_i(\textbf{x})=\sum_{j}{w_{ij}^- \; x_j} \: .
    \label{eq:d_x}
    \end{equation}
A matrix $\textbf{W}^+$ denotes only positive interactions, $w_{ij}^+=w_{ij}$ for $w_{ij}>0$. 
Otherwise, $w_{ij}^+ = 0$. 
Similarly, the elements of $\textbf{W}^- = [ w_{ij}^- ] $ are nonzero only when the interaction is negative.

For simplicity, we use the normalized abundance of species $i$ with respect to carrying capacity $K$, $f_i \equiv x_i/K$.
Then, Eq.~\eqref{eq:dxdt} is reduced as
    \begin{equation}
    \frac{df_i}{dt} = G_i(\textbf{f}) \; f_i \; \left( 1-\sum_j^{S(t)}{f_j} \right) \; + \; D_i(\textbf{f}) \; f_i \:,
    \label{eq:dfdt}
    \end{equation}
rescaling the time as $Kt \rightarrow t$ as the growth and death rates are rescaled.
Hereafter, we call this normalized abundance, $f_i$, as abundance for simplicity.

We assume that an invading species carries an initial abundance $f_0$.
Once new interactions are drawn between a new species and residents, the abundance dynamics follows Eq.~\eqref{eq:dfdt} with the given interaction structure $ [ w_{ij}^{\pm} ]$ before the next invasion event. 
We numerically integrate Eq.~\eqref{eq:dfdt} to get $f_i(t)$.
During integration, if the abundance $f_i$ becomes smaller than $1/K \equiv f_\text{th}$, we consider that species $i$ goes to extinction and set $f_i=0$. 
We use $f_0=10 \cdot f_\text{th}$ for simulations.

\begin{figure}
 \includegraphics[width=\linewidth]{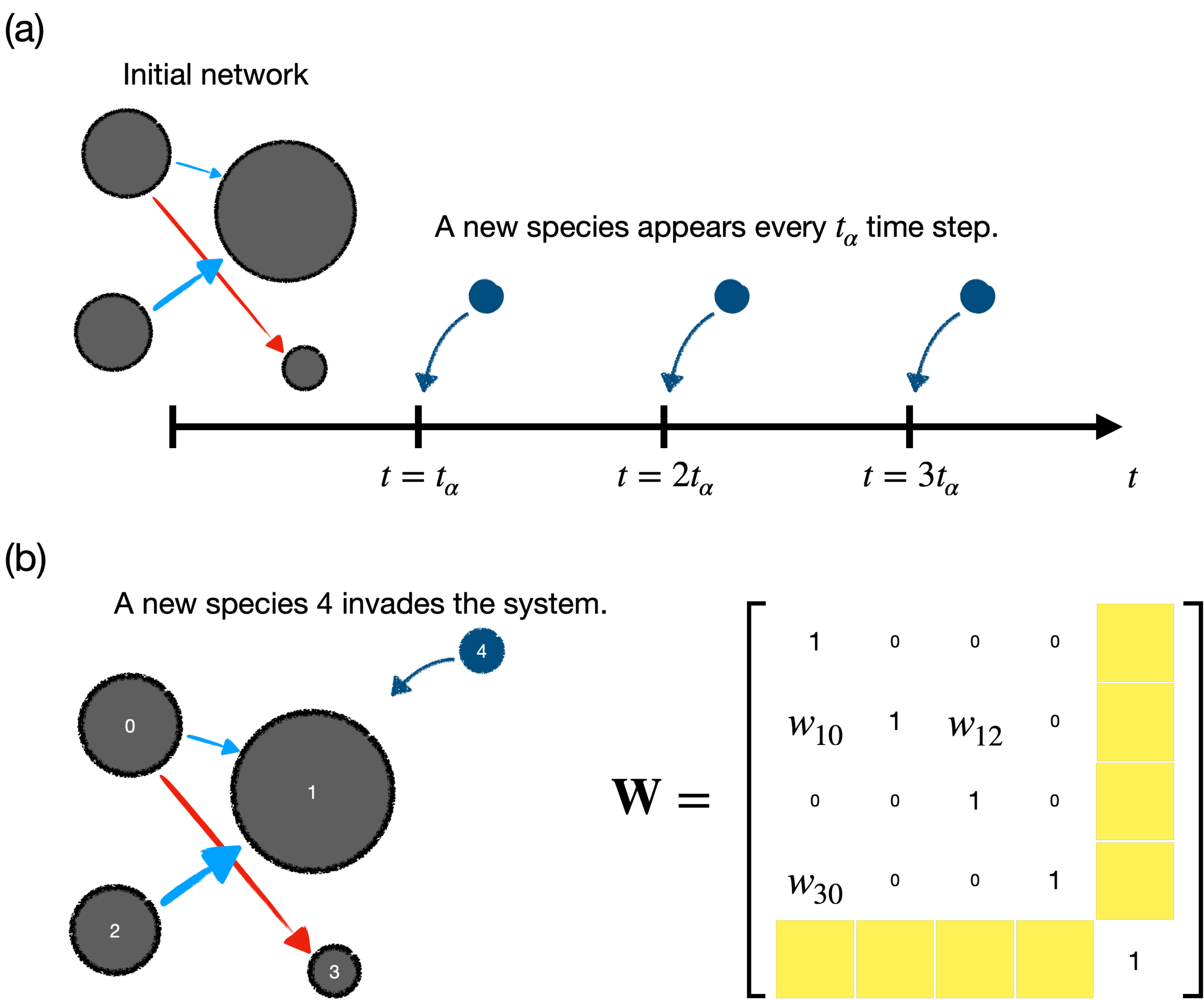}
 \centering
 \caption[Schematic figure for the model]{
 A schematic figure of an open evolving network model. 
 Species are represented by nodes and pair interactions are denoted by directional links with a sign.
 The direction of a link from $j$ to $i$ is drawn if node~(species) $j$ affects node (species) $i$.
 The positive and negative interactions are denoted by blue and red colors of the links.
 With those interactions, a generalized Lotka-Volterra type equation describes the abundance dynamics of species (see~Eq.~\eqref{eq:dfdt}).
 (a) Invasion events. A new species invades the system every $t_{\alpha}$ time step, where the interval between two invasions is $t_\alpha \equiv K/\alpha$. 
 (b) New interactions. A new species interacts with randomly chosen $m$ resident species.
 We set $m=5$ because the previous study with a simpler dynamical rule showed that the system can achieve high diversity under the same ``open and evolving'' condition for $4 < m < 19$~\cite{shimada2014universal}.
 Interaction weight $w_{ij}$ from species $j$ to $i$ is sampled from the normal distribution with zero mean and standard deviation $\sigma$.
 If there are no interactions from $j$ to $i$, weight $w_{ij}$ is zero.
 We set intraspecific interaction weight $w_{ii}$ as unity because isolated species can survive by themselves by taking resources.  
}
 \label{fig:scheme}
\end{figure}

\begin{figure}
 \centering
 \includegraphics[width=0.84\linewidth]{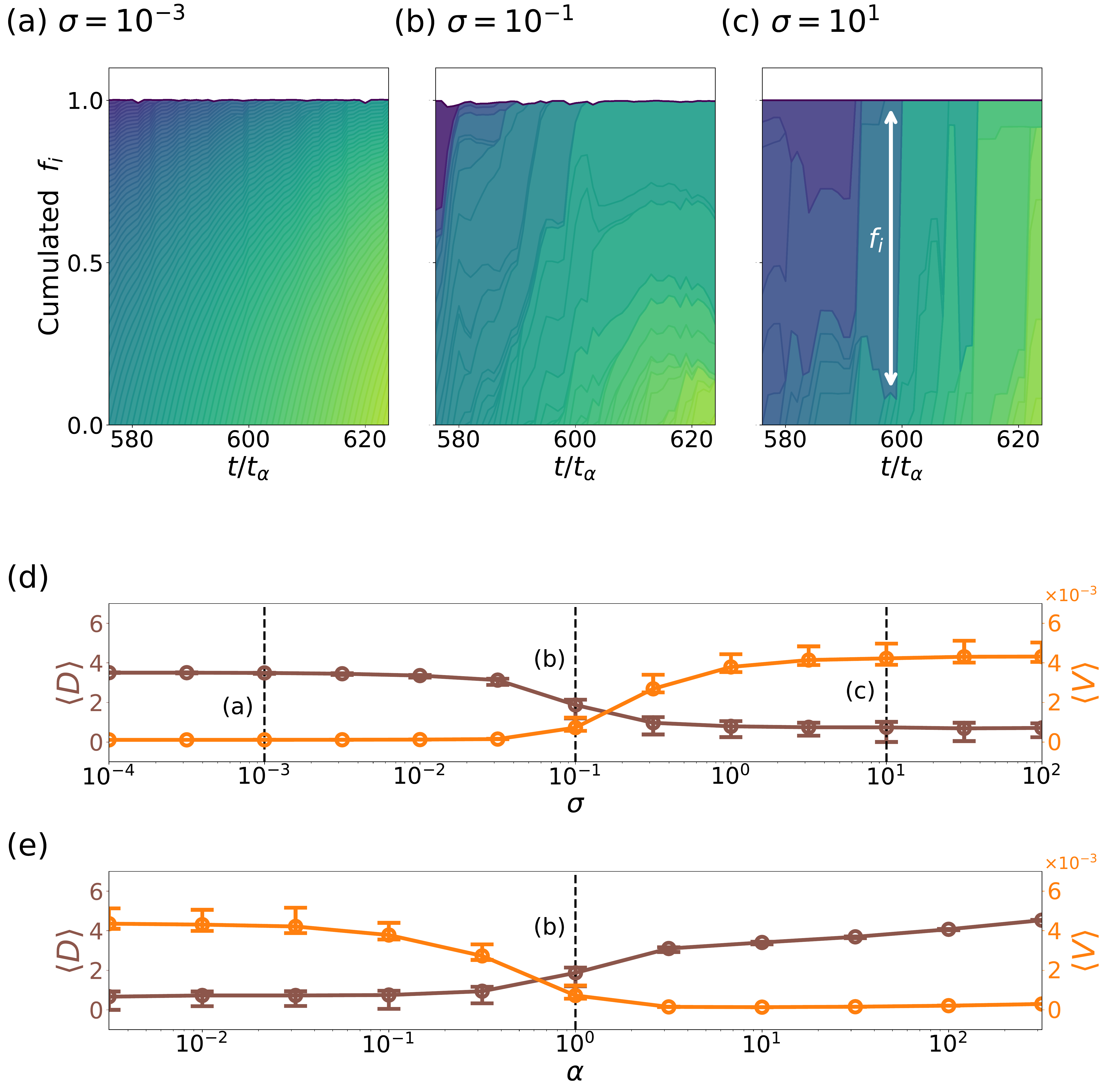}
 \caption[Abundance dynamics according to $\sigma$]{
 Abundance dynamics (a)-(c) for the given invasion rate~$\alpha=1$ and various interaction strengths~$\sigma=10^{-3}$,~$10^{-1}$, and~$10^{1}$.
 Using different colors for different species introduced at different times, we stack abundance $f_i$ so that the height of each color indicates the abundance of each species at time $t/t_{\alpha}$.
 The width indicates the lifetime of a species.
 Colors indicate the age of each species. 
 Blue and yellow are for the oldest and newest species, respectively. 
 For better visualization, we show the abundance dynamics only for $t/t_{\alpha} \in [580,620]$. (d)-(e) Shannon's diversity index~$D$ and variability~$V$ are measured in the steady state for 100 independent simulations.
 The error bars indicate a 95\% confidence interval.
 The dashed lines indicate the parameter sets used in (a)-(c), respectively.
 Increasing interaction strength~$\sigma$ gives a similar result with decreasing invasion rate~$\alpha$. 
 We used the total carrying capacity~$K=100$ for the visualization (the trends in (d)-(e) are robust for larger $K$).
 Note that we also measure various population-level quantities in the steady state (see Supplementary~Information~B for detail).
 }
 \label{fig:preview}
\end{figure}

\subsection{Evolving interaction network} \label{subsec:network}
If we treat species as nodes, the interaction $w_{ij}$ can be mapped into a link from $j$ to $i$ with a direction and a sign.
Thus, the interaction structure can be represented by a network. 
This network structure changes over time due to invasions and extinctions of species. 
We suppose that a new species comes into the system with the invasion rate $\alpha$. 
It means that a new species appears in the system every $t_\alpha~\equiv~K/\alpha$ on average. 
The smaller $\alpha$ is, the less frequent new species come into the system.
For simplicity, we add a new species to the system every $t_\alpha$~(see~Fig.~\ref{fig:scheme}(a)).

When an invasion event happens, the number of nodes in the network increases by one, and $m$ new interactions with this new species appear (see~Fig.~\ref{fig:scheme}(b)).
To assign the new interaction weights, we sample the weights $w_{ij}$ from the normal distribution $\mathcal{N}(0,\sigma^2)$, except for the intraspecific interaction weight ($w_{ii}=1$).
That is, the larger $\sigma$ is, the stronger the interactions become.
The sign of $w_{ij}$ indicates the effect of species $j$ on the abundance of species $i$.
After an invasion event, the system follows the dynamics as described in Eq.~\eqref{eq:dfdt}.
When species go extinct, the nodes and the links attached to those nodes are removed.

We initially construct a random network with $S_0$ nodes and $m$ links for each node, and randomly assign the direction and the weight of all links.
To investigate how invasion and interaction shape the population composition, we constantly add a new species until the system has a saturated number of surviving species. The number of nodes is denoted by $S(t)$.
We call this regime \emph{steady} when $\left[S(t+\Delta t) - S(t)\right] / \Delta t \sim 0$ and observe how the abundance of surviving species changes over time depending on invasion rate $\alpha$ and interaction strength $\sigma$ (see Supplementary~Information~A for detail).

\begin{figure*}
 \centering
 \includegraphics[width=\linewidth]{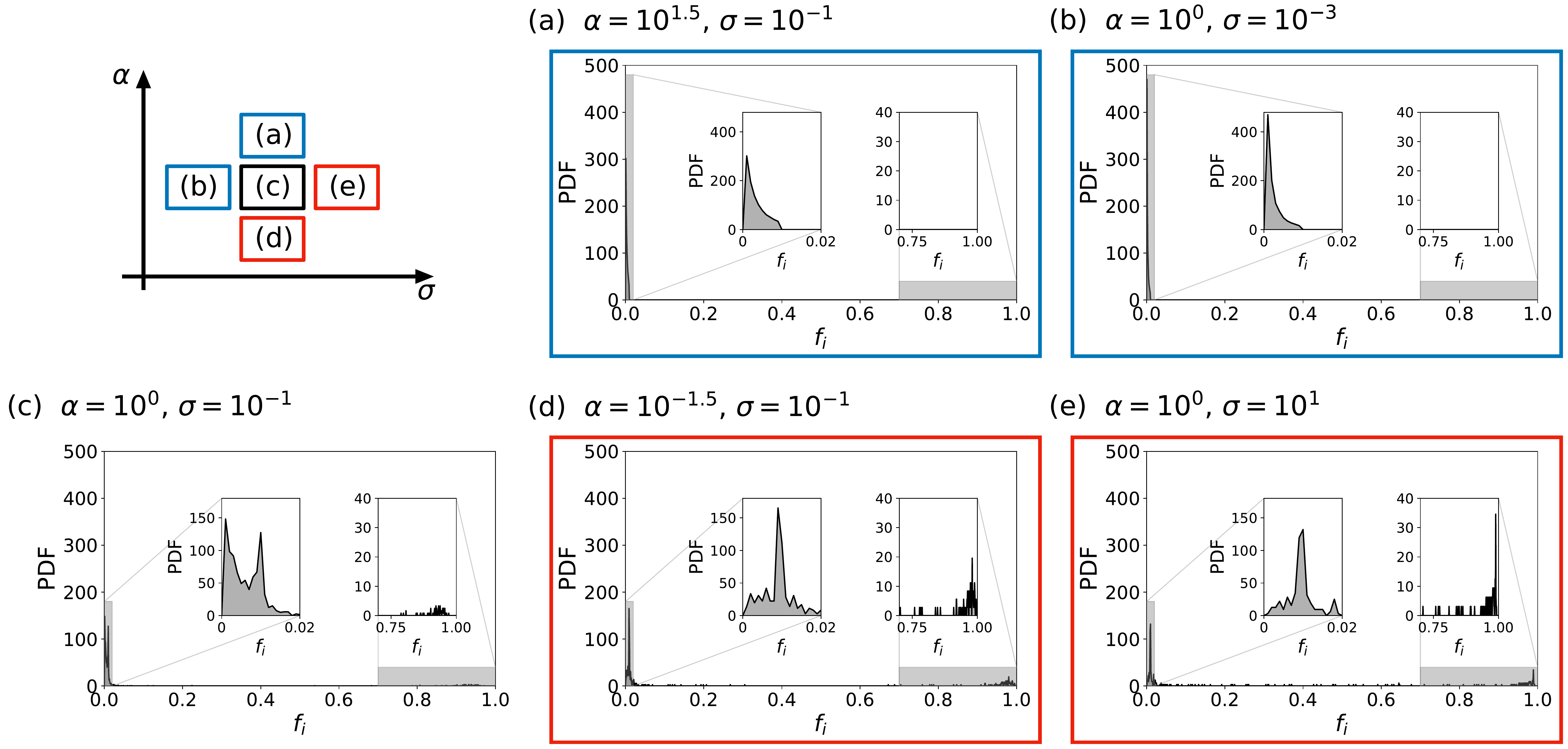}
 \caption[Abundance distribution]{
 Abundance distributions at the steady state for various invasion rates~$\alpha=10^{1.5}$,~$10^{0}$, and~$10^{-1.5}$ and interaction strengths~$\sigma=10^{-3}$,~$10^{-1}$,and~$10^{1}$.
 Ranges $f_i\in[0,0.02]$ and $f_i\in[0.7,1]$ are zoomed in the insets to check the distribution shapes in very small and large abundances.
 If the invasion occurs too frequently to reach the equilibrium of the system, the distribution will have no dominant species.
 On the other hand, when the interaction strength is strong enough to reach equilibrium within two successive invasion events, dominant species can emerge.
 The color of a panel frame indicates which effect dominates at the given parameter set.
 The blue color means the invasion is dominant and the red one indicates the interaction is dominant in the system.
 We used the total carrying capacity~$K=1000$.
 For obtaining the distributions, $100$ independent simulations are used.
 }
 \label{fig:distributions}
\end{figure*}

\section{Result} \label{sec:result}
Depending on interaction strength~$\sigma$, abundance dynamics shows different behaviors in the steady state at a given invasion rate $\alpha$ (see~Figs.~\ref{fig:preview}(a)-(c)). 
In a weak-interaction limit among species, the invasion rate solely determines the whole abundance.
In this case, existing species are gradually pushed out of the system due to constant invasions. 
A similar scenario happens for low $\sigma$ where the invasion dominates interactions (see~Fig.~\ref{fig:preview}(a)). 
In contrast, for large $\sigma$, the interactions are strong enough to induce the species with large abundance, reducing the diversity of the system (see Figs.~\ref{fig:preview}(b) and \ref{fig:preview}(c)).
At the same time, new interactions introduced by an invader are also strong enough to kill such dominant species.
Thus, the dominant species keep changing over time (see~Fig.~\ref{fig:preview}(c)).

To quantify the decrease in diversity for strong interactions, we measure Shannon's diversity index~$D$ for various $\sigma$ as
    \begin{equation}
    D=-\sum_{i}{f_i \log{f_i}} \:.
    \label{eq:diversity}
    \end{equation}
Another feature of the abundance dynamics for strong interactions is fast changes in dominant species (see~Fig.~\ref{fig:preview}(c)).
As a new dominant species emerges all the time, the fluctuation of the abundance is large.     
To quantify changes in abundance over time, we measure variability~$V$ as follows:
    \begin{equation}
    V=\Big< {\:\text{Var}_t(f_i)} \Big>_S \:,
    \label{eq:variability}
    \end{equation}
where $\text{Var}_t(f_i)$ indicates the variance of abundance $f_i$ over time and $\big< \cdot \big>_S$ means average over the number of species that survived at the steady state.
The large variability $V$ indicates frequent turnover of dominant species.
As increasing interaction strength~$\sigma$, diversity decreases while variability becomes higher at a fixed invasion rate (see~Fig.~\ref{fig:preview}(d)).

Even though the interaction strength $\sigma$ is large, the system can have high diversity and low variability when the invasion rate $\alpha$ is large enough.
It is because the species expected to be dominant is pushed out before reaching its equilibrium due to the high invasion rate.
Thus, increasing $\alpha$ has the opposite effect of increasing $\sigma$ (see~Fig.~\ref{fig:preview}(e)).
As a high invasion rate can inhibit the effect of interactions, the interplay between invasion rate $\alpha$ and interaction strength $\sigma$ determines the abundance dynamics.

The species with small abundance (around the initial abundance) result from invasions, while strong interaction is essential to observe the species with large abundance (much larger than the initial abundance $f_0$).
If the invasion effect is dominant, all abundances will be around initial values.
On the other hand, species with a large abundance can appear for higher interaction strength.
Figure~\ref{fig:distributions} shows the steady-state abundance distributions for various invasion rates $\alpha$ and interaction strength $\sigma$.
Increasing $\alpha$ and decreasing $\sigma$ give the same effects on abundance distributions (see~Figs.~\ref{fig:distributions}(a)~and~\ref{fig:distributions}(b)).
In the same way, decreasing $\alpha$ and increasing $\sigma$ show the similar trends of abundance distributions (see~Figs.~\ref{fig:distributions}(d)~and~\ref{fig:distributions}(e)).
Invasion rate~$\alpha$ and interaction strength~$\sigma$ have the opposite effects on abundance distributions.

When the invasion events happen too frequently, no dominant species appear (see~Figs.~\ref{fig:distributions}(a)-(b)).
In this regime, species have abundances less than or equal to the initial value ($f_i \leq f_0$ in the figure).
Once a new species comes into the system, all abundances are reduced proportionally to their abundances due to resource limitation.
Thus, a portion of each abundance remains in the system. 
A newly added species undergoes such a reduction once, while the residents already experienced several dilutions. 
Thus the later invading species have the larger abundance, showing decreasing distribution function of~$f_i$.

Conversely, in the interaction dominant regime (see~Figs.~\ref{fig:distributions}(d)-(e)), dominant species appear with non-zero probability.
Species with small abundances can hardly survive except for new invading species. 
That is the reason why the peak around $f_i=f_0$ remains.

\begin{figure}
 \centering
 \includegraphics[width=\linewidth]{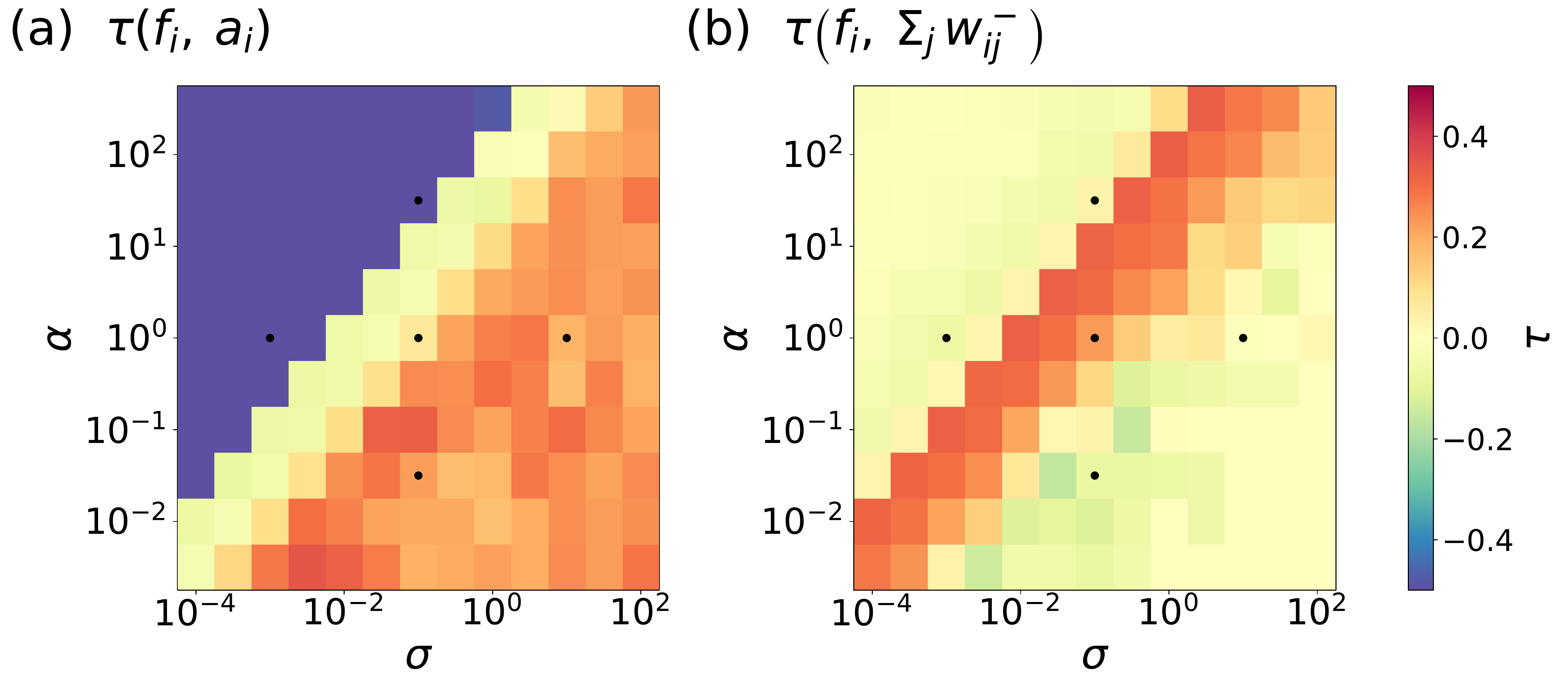}
 \caption[Correlations]{
 Kendall's tau-b correlation coefficients $\tau$ between abundance $f_i$ and age $a_i$~in (a), and the correlation coefficients between the abundance $f_i$ and the negative incoming weight sum $\sum_j\: w_{ij}^-$ in (b), respectively.
 The correlation coefficients are measured at the steady state.
 In the invasion dominant regime (upper left), the abundance negatively correlates with its age because each invasion event pushes down all resident species abundances. 
 Moreover, the interaction weights are not correlated with the abundance because the interaction effect is not enough to change the observed abundance dynamics.
 In the lower right of the panels, however, the interaction effect is much more dominant than the invasion effect on the dynamics.
 Once an invading species has lots of positive links from others, the species have a higher abundance.
 Then, the species dominantly survive until new dominant species appear while other invading species with not enough positive links are pushed out immediately.
 For this reason, the abundance of species positively correlates with the age in the interaction dominant regime because we measure the correlation coefficient for surviving species at the steady state.
 In the lower right in (b), the interaction-dominant regime, species with negative incoming links cannot survive.
 As a result, the correlation between abundances and negative incoming links cannot be captured as the surviving species do not have negative incoming links (see~Fig.~S2(g)).
 Black dots correspond to the parameter sets used in Fig.~\ref{fig:distributions}.
 The total carrying capacity is set to~$K=1000$ (see Supplementary~Information~C for the details).
 }
 \label{fig:corr}
\end{figure}

To examine how invasion rate~$\alpha$ and interaction strength~$\sigma$ determine the patterns of the observed abundance dynamics, we imagine two extreme cases: one is the case of zero interactions given at a constant invasion rate, and the other is strong interactions that ensure the equilibration of the system before new invasions.

Without interactions ($\sigma=0$), each species is independent and identical.
There are no differences between species except for their age--how long the species stay in the community. 
Once a new invasion event occurs, this invader pushes down all residents' abundances due to the limited population size (see~Eq.~\eqref{eq:dfdt}).
Thus, elder species that undergo more invasion events are likely to have a smaller abundance compared to others, giving a smooth gradient of abundances between species.
Therefore, negative correlations between abundances and ages appear for large $\alpha$ and small $\sigma$ (see~Fig.~\ref{fig:corr}(a)).
On the contrary, young species tend to have smaller abundances for small $\alpha$ and large $\sigma$ as dominant species emerge and invaders usually die out.
 
Furthermore, we find that species abundances are highly correlated with the sum of negative incoming weights only in the intermediate regime between invasion-dominant and interaction-dominant regimes. 
The positive correlations between abundances and the sum of negative incoming weights in the interaction-dominant regime are expected because the derivatives of abundances are highly dependent on their interaction structure at the steady-state~(see~Eq.~\eqref{eq:dfdt}). 
However, too strong an interaction leads to almost immediate deaths of the other species, and all observed species and interactions are transient~(see~Fig.~S2(g)).
The calculated correlation coefficients for various $\alpha$ and $\sigma$ support that there are two regimes where either invasion or interaction dominates. 

\section{Discussion} \label{sec:discussion}
Interacting species can hardly form a high-diversity community because the fittest species takes all resources.
However, with a high invasion rate, ecosystems can show high diversity.
When invasion events occur before the system reaches its equilibrium, multiple species can coexist.
In other words, the interaction strength controls the equilibrium time of the dynamics, while the invasion rate interrupts reaching equilibrium by changing the interaction structure.
Therefore, invasion and interaction play an important role in shaping the population composition (see Supplementary~Information~D for detailed calculation).

To describe the dynamics according to the invasion rate $\alpha$ and the interaction strength $\sigma$, we consider the Lotka-Volterra type equation with an open evolving interaction network (see~Eq.~\eqref{eq:dfdt}).
As a new invader enters the system with a fixed time interval, new interactions between the invader and the residents are drawn.
Because the system undergoes invasions and extinction of species, the growth and death rates of all species change over time.
We found that the invasion rate $\alpha$ and the interaction strength $\sigma$ have opposite effects on the population composition. 
The system has a high diversity and low variability for high invasion rate~$\alpha$ and small interaction strength~$\sigma$.
In contrast, when the interaction strength $\sigma$ is sufficiently strong, dominant species appear. 
The composition is almost homogeneous when the invasion effect is dominant, while the interaction effect makes the population composition become heterogeneous.

Our finding shows that fast invasion events not only provide temporal coexisting species with small abundances but shift the regime, governing the observed abundance dynamics with interaction strength. 
Thus even if strong competition tends to result in the emergence of dominant species, fast invasion events prevent the growth of species which is expected to be dominant.
It means that not only the interaction strength but also the invasion rate is important in shaping the population composition. 
It is because invasions introduce a new time scale in the system, the invasion time scale. 

\section*{acknowledgments} \label{sec:acknowledgments}
Y. Park and H.J. Park were supported by the National Research Foundation (NRF) grant funded by the Korea Government (MSIT) Grant No.2020R1A2C1101894. 
S.-W. Son was supported by the NRF of Korea through Grant No. NRF-2020R1A2C2010875 and also partially supported by the Institute of Information \& communications Technology Planning \& Evaluation (IITP) grant, funded by the MSIT, No.2020-0-01343, Artificial Intelligence Convergence Research Center (Hanyang University ERICA). 
This work was supported by INHA UNIVERSITY Research Grant as well. We also acknowledge the hospitality at APCTP where part of this work was done.
This work was supported under the framework of international cooperation program managed by the National Research Foundation of Korea (NRF-2022K2A9A2A07000211).

\section*{Data availability} \label{sec:data}
The code about the open evolving network model is uploaded at \href{https://github.com/youngjai/OpenEvolvingSystem}{https://github.com/youngjai/OpenEvolvingSystem}.

\section*{References}
\bibliography{bibliography}

\end{document}


\author{Youngjai Park}
 \affiliation{Department of Applied Physics, Hanyang University, Ansan 15588, Korea}
 \affiliation{Asia Pacific Center for Theoretical Physics (APCTP), Pohang 37673, Korea}

\author{Takashi Shimada}%
 \affiliation{Mathematics and Informatics Center and Department of Systems Innovation, The University of Tokyo, Tokyo 113-8656, Japan}
 
\author{Seung-Woo Son}
 \email{sonswoo@hanyang.ac.kr}
 \affiliation{Department of Applied Physics, Hanyang University, Ansan 15588, Korea}
 \affiliation{Asia Pacific Center for Theoretical Physics (APCTP), Pohang 37673, Korea}
 \affiliation{Center for Bionano Intelligence Education and Research, Hanyang University, Ansan, 15588, Korea}

\author{Hye Jin Park}
 \email{hyejin.park@apctp.org}
 \affiliation{Asia Pacific Center for Theoretical Physics (APCTP), Pohang 37673, Korea}
 \affiliation{Department of Physics, POSTECH, Pohang, 37673, Korea}

\begin{center}
\Large{Supplemental Information (SI)\\ for ``Invasion and Interaction Determine Population Composition in an Open Evolving System"}
\end{center}
\begin{center}
Youngjai Park, Takashi Shimada, Seung-Woo Son, Hye Jin Park
\end{center}
\tableofcontents

\appendix

\addcontentsline{toc}{section}{Appendices}
\counterwithout{equation}{section}
\renewcommand{\thefigure}{S\arabic{figure}}
\renewcommand{\theequation}{S\arabic{equation}}
\renewcommand{\thetable}{S\arabic{table}}

\newpage
\section{Steady state} \label{app:steady}

\begin{figure}[h]
 \includegraphics[width=0.9\linewidth]{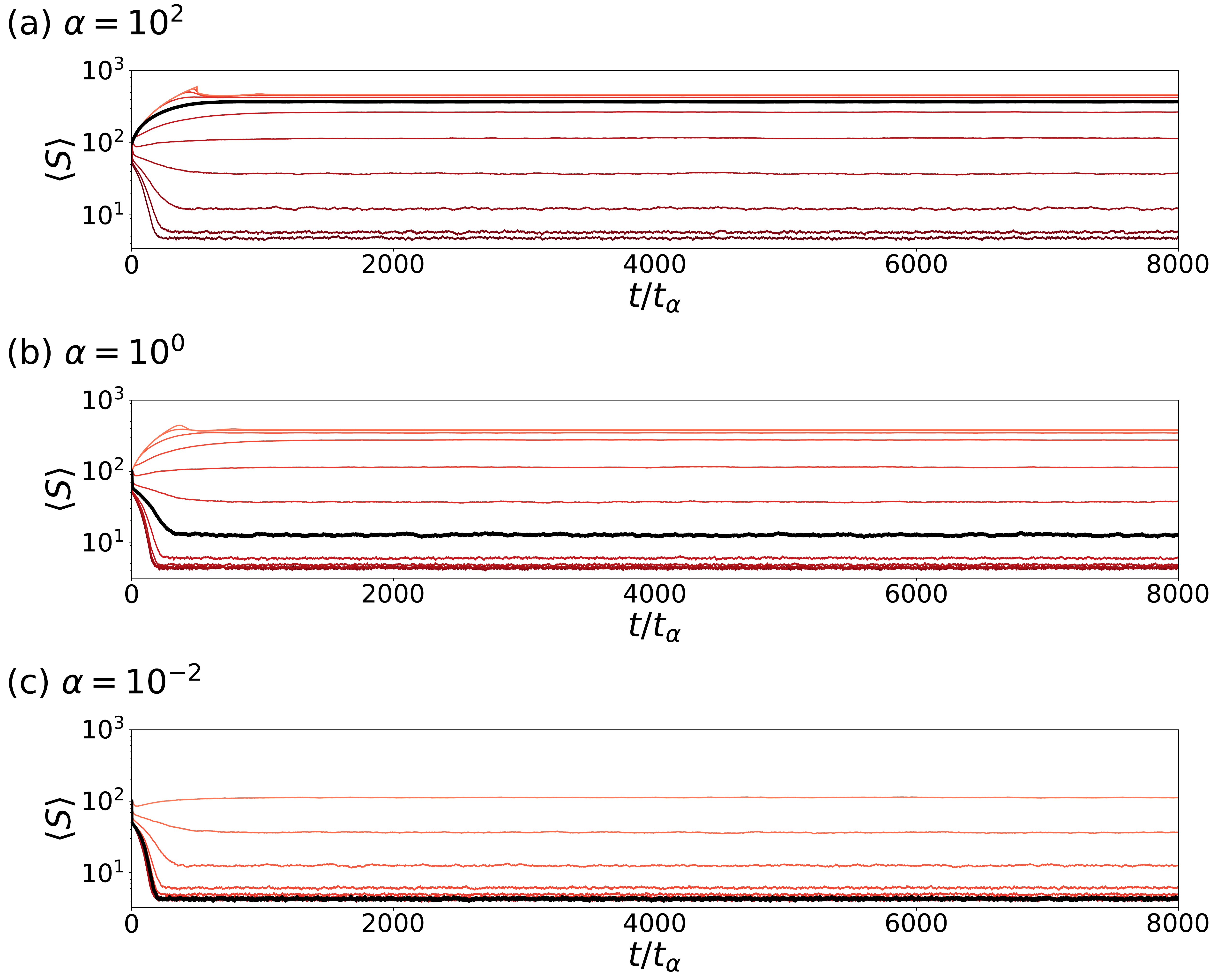}
 \centering
 \caption[Steady state]{
 Richness~$\langle S \rangle$, which means the number of surviving species, over the invasion event time with the carrying capacity~$K=1000$.
 The colors of the lines indicate interaction strength. 
 According to interaction strength $\sigma$, we draw the lines from dark red (low $\sigma$) to bright red (high $\sigma$).
 The black lines mean~$\sigma=10^{-1}$. 
 All lines show the $100$ ensemble-averaged values.
 }
 \label{fig:steady-state}
\end{figure}

In the open evolving network model, a new species appears at every given interval.
In this sense, the invasion plays a perturbation role in the system.
To quantitatively examine the system, we need to confirm when the system reaches a steady state.
Figure~\ref{fig:steady-state} shows the change of the number of surviving species over time for given invasion rate~$\alpha$ and interaction strength~$\sigma$.
In the main text, we consider the steady-state at $t/t_\alpha = 6000$.

\newpage
\section{Statistical properties} \label{app:statistics}

\begin{figure}[h]
 \includegraphics[width=0.9\linewidth]{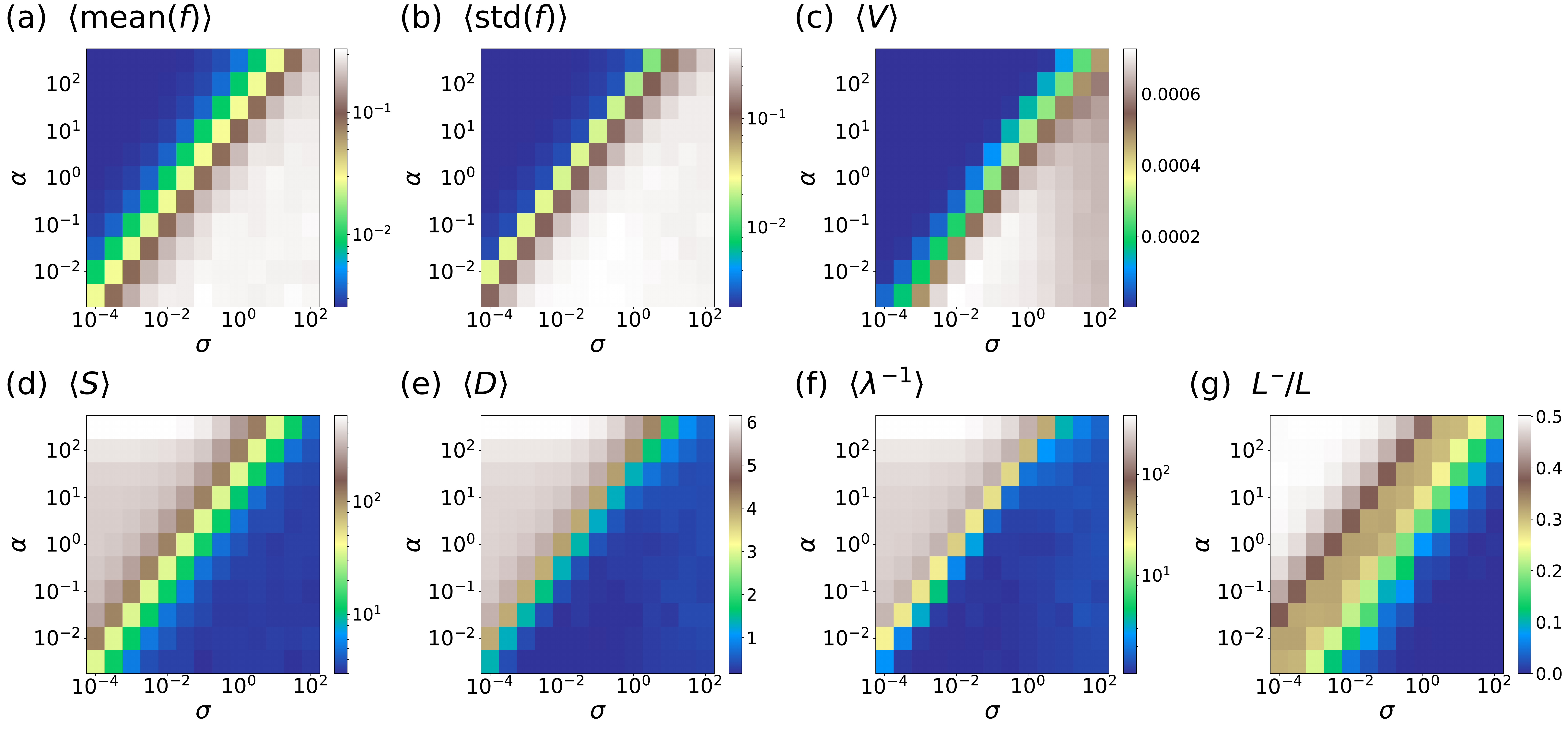}
 \centering
 \caption[Diversity indexes]{
 The mean and standard deviation of abundances (a)-(b) at the given time $t/t_{\alpha}=6000$, and the variability~$V$ (c) for the range from $5000$ to $7000$.
 Richness~$S$, Shannon's index~$D$, the inverse Simpson's index~$\lambda^{\text{-}1}$, and the fraction of the number of negative links~${L^\text{--}/L}$ (d)-(g) at the given time $t/t_{\alpha}=6000$.
 We show the heatmaps according to the invasion rate~$\alpha$ and the interaction strength~$\sigma$ with $K=1000$.
 Each value of sites in the figures presents the average using $100$ independent realizations.
 }
 \label{fig:properties}
\end{figure}

To investigate how the system works, we measure several statistical properties in invasion rate~$\alpha$ and interaction strength~$\sigma$ space.
Generally, we can present the diversity index in three ways.
The simplest way is richness~$S$ which is just the number of surviving species~\cite{tuomisto2010consistent}.
If we consider the abundance distribution as well as the richness, we can take Shannon's entropy~$D=-\sum {f_i \log{f_i}}$, which usually applies in the field of information theory~\cite{shannon1948mathematical,spellerberg2003tribute,jost2006entropy}.
The other way is the inverse Simpson's diversity~$\lambda^{\text{-}1}=1/\sum{{f_i}^2}$, which indicates the effective number of surviving species considering the abundance distribution and richness~\cite{jost2006entropy}.

\newpage
\section{Correlations} \label{app:correlations}

\begin{figure}[h]
 \includegraphics[width=0.9\linewidth]{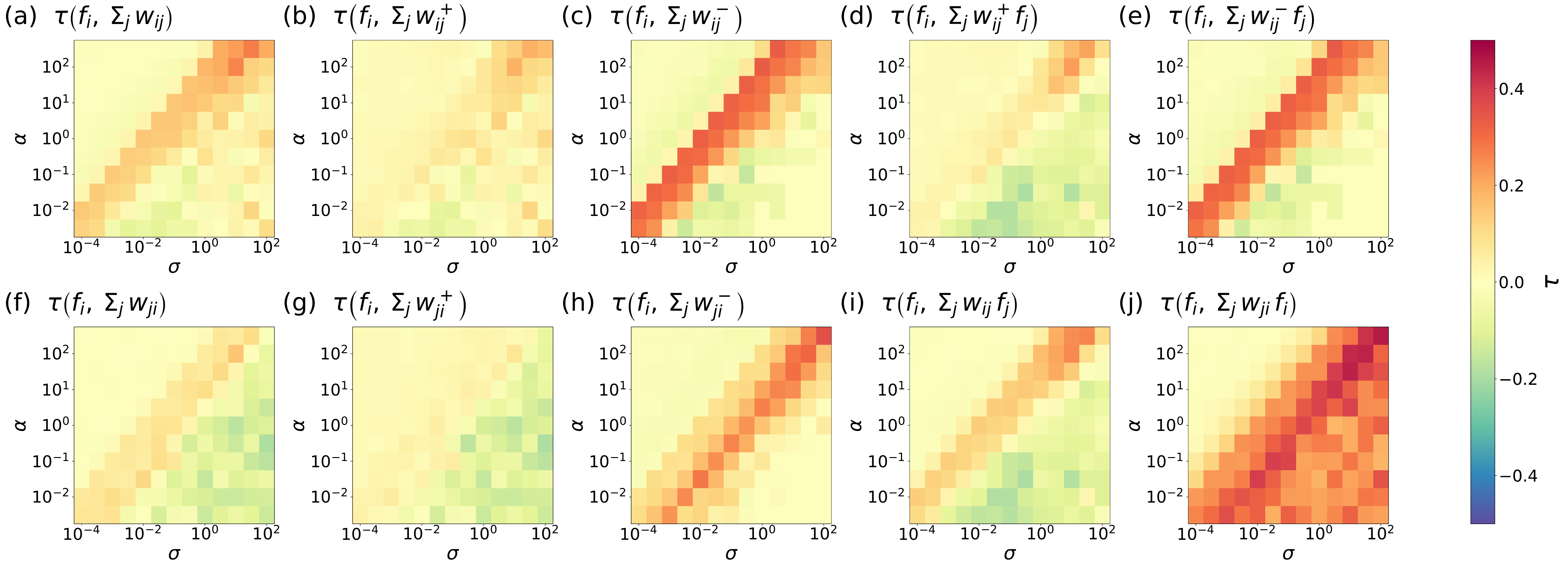}
 \centering
 \caption[Correlation coefficients]{
 Kendall's tau-b correlation coefficients between the abundance and the interaction structures at time $t/t_\alpha=6000$.
 We show the heatmaps according to the invasion rate~$\alpha$ and the interaction strength~$\sigma$ with $K=1000$.
 Each value presents the average by gathering $100$ independent configurations.
 }
 \label{fig:corr_all}
\end{figure}

The model has a removal and invasion process over time.
Both the invasion and the interaction affect the correlation coefficients between the abundance and the properties related to the network structure.
Figure~\ref{fig:corr_all} shows Kendall's tau-b correlation coefficient between the abundances and these network properties~\cite{agresti2010analysis}.
For example, we notate the coefficient between two series, $\textbf{x}$ and $\textbf{y}$, as~$\tau\left(\textbf{x},\textbf{y}\right)$.
The properties of the positive weights do not seem to correlate with the abundance. 
However, the ones related to negative weights show non-zero coefficients at the intermediate region in invasion rate~$\alpha$ and interaction strength~$\sigma$ space.

\newpage
\section{Dynamics without invasion}
\label{app:no_invasion}

As we see in the main text, one feature of our suggested model is a dominant species appears in the interaction dominant regime (see~Figs.~2~and~3).
To analyze the mechanism of why the dominant species can appear in the regime, we first try to understand the Lotka-Volterra-type dynamics without the invasion process.
The model follows as
    \begin{equation}
    \frac{df_i}{dt} = G_i(\textbf{f}) \; f_i \; \left( 1-\sum_j^{S}{f_j} \right) \; + \; D_i(\textbf{f}) \; f_i \:,
    \label{eq:dfdt_1}
    \end{equation}
with $G_i(\textbf{x})=\sum_{j}{w_{ij}^+ \; x_j}$ and $D_i(\textbf{x})=\sum_{j}{w_{ij}^- \; x_j}$ (see~Eq.(5)).

To interpret the dynamics, we consider $F \equiv \sum_i^S {f_i}$ and $q_i \equiv f_i/F$ with constraint condition $\sum_i^S {q_i} = 1$.
Then, the dynamics of $F$ follow as
    \begin{equation}
    \begin{split}
    \frac{dF}{dt} & = \sum_i^S {\left[
        \left(\sum_{j\setminus i}{w_{ij}^+f_j + f_i}\right)\:
        f_i \: 
        \left( 1-\sum_k^{S}{f_k} \right) 
        +\sum_{j\setminus i}{w_{ij}^-f_jf_i}
        \right]} \\
    & = F^2\sum_i^S{\left[\left(\sum_{j\setminus i}{w_{ij}^+q_jq_i}+{q_i}^2 \right)(1-F) +\sum_{j\setminus i}{w_{ij}^-q_jq_i} \right]} \\
    & = F^2 \left[
        \left(\sum_i^S{{q_i}^2}+\sum_i^S\sum_{j\setminus i}{w_{ij}^+q_jq_i}\right)(1-F)
        +\sum_i^S\sum_{j\setminus i}{w_{ij}^-q_jq_i} \right] \:,
    \end{split}
    \label{eq:dFdt}
    \end{equation}
and the dynamics of frequency $q_i$ follows
    \begin{equation}
    \frac{dq_i}{dt} = \frac{d}{dt}\left(\frac{f_i}{F}\right) 
    = \frac{1}{F^2}\left(\frac{df_i}{dt}F-f_i\frac{dF}{dt}\right) 
    = \frac{1}{F}\left(\frac{df_i}{dt}-q_i\frac{dF}{dt}\right)\:.
    \label{eq:dqdt_chain}
    \end{equation}
The first term of RHS in Eq.\eqref{eq:dqdt_chain} is
    \begin{equation}
    \begin{split}
    \frac{1}{F}\frac{df_i}{dt} & = \frac{1}{F}
        \left[\left(\sum_{j\setminus i}{w_{ij}^+f_j + f_i}\right)\: 
        f_i \:
        \left( 1-\sum_k^{S}{f_k} \right) 
        +\sum_{j\setminus i}{w_{ij}^-f_jf_i}\right] \\
    & = F\left[
        \left(\sum_{j\setminus i}{w_{ij}^+q_j + q_i}\right) 
        q_i(1-F) 
        +\sum_{j\setminus i}{w_{ij}^-q_jq_i}\right] \:.
    \end{split}
    \label{eq:dqdt_1}
    \end{equation}
Thus,
    \begin{multline}
    \frac{dq_i}{dt} = Fq_i \Bigg[ \Bigg. 
        \left\{\left(\sum_{j\setminus i}{w_{ij}^+q_j + q_i}\right)(1-F)+\sum_{j\setminus i}{w_{ij}^-q_j}\right\} \\
        -\left\{\left(\sum_i^S{{q_i}^2}+\sum_i\sum_{j\setminus i}{w_{ij}^+q_jq_i}\right)(1-F)
        +\sum_i^S\sum_{j\setminus i}{w_{ij}^-q_jq_i}\right\} 
        \Bigg. \Bigg] \:,
    \label{eq:dqdt_2}
    \end{multline}
Therefore, $\dot{F}$ and $\dot{q_i}$ follow as
    \begin{equation}
    \begin{split}
    & \frac{dF}{dt} = F^2 \left[
        \left(Q+\left\langle g_i(\textbf{q})\right\rangle_{\textbf{q}}\right)(1-F)
        +\left\langle D_i(\textbf{q})\right\rangle_{\textbf{q}}
        \right] \:, \\
    & \frac{dq_i}{dt} = Fq_i \left[ 
        \left(q_i+g_i(\textbf{q})-Q-\left\langle g_i(\textbf{q}) \right\rangle_{\textbf{q}}\right)(1-F)
        +D_i(\textbf{q})-\left\langle D_i(\textbf{q}) \right\rangle_{\textbf{q}}
        \right] \:,
    \end{split}
    \label{eq:dFdt_dqdt}
    \end{equation}
where $Q \equiv \sum_i^S{{q_i}^2}$ (we already measured $Q$ as $\lambda^{-1}$, $Q \equiv F^{-2}\lambda$, in Fig.\ref{fig:properties}.) and $g_i(\textbf{q})$ means the sum of the frequency of species $i$ with its positive weight except for itself, $g_i(\textbf{q}) = G_i(\textbf{q}) - q_i = \sum_{j\setminus i}{w_{ij}^+q_j}$ and $\langle \cdots \rangle_{\textbf{q}}$ indicates an average according to vector $\textbf{q}$.

\subsection{Without interactions ($\sigma=0$)}
\label{subapp:wo_interactions}

\begin{figure}[h]
 \includegraphics[width=0.4\linewidth]{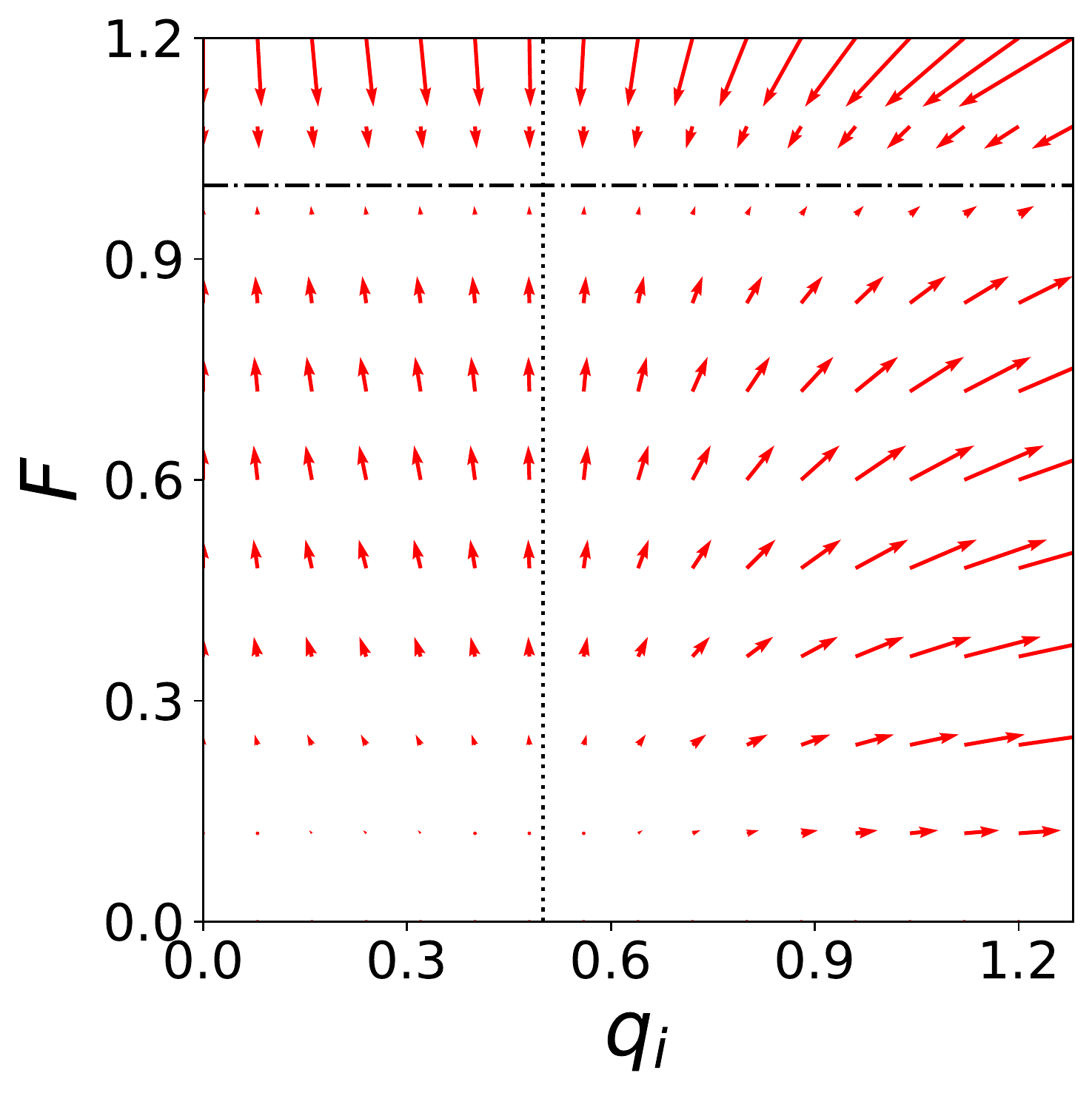}
 \centering
 \caption[Phase portrait]{
 Phase portrait, $q_i$ versus $F$. We represent the sum of the normalized abundance as $F \equiv \sum_i^S{f_i}$ and the frequency of specie $i$ as $q_i \equiv f_i/F$. Arrows show the direction depending on the derivative values and the width of the arrow indicates the magnitude of the values. The dash-dot line, $F=1$, indicates nullcline and the dotted line means $Q$, which we set the value as a constant, $Q=0.5$, to intuitively visualize the portrait.
 }
 \label{fig:portrait}
\end{figure}

First, we consider that there are no interactions between species for simplicity.
Then, Eq.\eqref{eq:dFdt_dqdt}~is
    \begin{equation}
    \begin{split}
    & \frac{dF}{dt} = F^2 (1-F) Q \:, \\
    & \frac{dq_i}{dt} = F (1-F) q_i \left(q_i-Q\right) \:.
    \end{split}
    \label{eq:dFdt_dqdt_non}
    \end{equation}
In this case, the stable fixed point of $F^*$ is $F^*=1$ due to $Q \geq 0$.
For $q_i$, we consider three cases.
    \begin{equation}
    \begin{split}
        & \text{(i)}\; F > 1 : {q_i}^* = Q \:. \\
        & \text{(ii)}\; F = 1 : {q_i}^* = {q_i}_{t_{st}} \:. \\
        & \text{(iii)}\; F < 1 : \\
        & \qquad\qquad \text{(a)}\; {q_i}_0 > Q : {q_i}^* \rightarrow \infty \:, \\
        & \qquad\qquad \text{(b)}\; {q_i}^* = Q : {q_i}^* = 0 \:.
    \end{split}
    \label{eq:cases}        
    \end{equation}
If we consider the initial abundance $q_{i0}$ for all $i$ to be small enough, we can assume that $F$ starts from a value less than 1.
In this case, $q_i$ survives (or~extincts) according to its initial abundance is larger (or~smaller) than $Q$ because the dynamics will follow case (iii).
Additionally, $q_i$ can't be over $F$ because of the constraint $\sum_i^S{q_i}=1$.

The dynamics of $Q$ is as follows: 
    \begin{equation}
    \frac{dQ}{dt} = \frac{d}{dt}\left(\sum_i^S{{q_i}^2}\right) = \sum_i^S{\left(2q_i\frac{dq_i}{dt}\right)}
        = 2\sum_i^S{\dot{q_i}q_i} = 2 \left\langle \dot{q_i} \right\rangle_{\textbf{q}}\:.
    \label{eq:dQdt}
    \end{equation}
That is, $\dot{Q}$ means the weighted average of $\dot{q_i}$ with vector $\textbf{q}$. 
When the dynamics starts from $F<1$, which we assume, the larger $q_i$ is, the larger $\dot{q_i}$ is also.
As constraints $\sum_i^S{q_i}=1$ and $\sum_i^S{\dot{q_i}}=0$, we can show that $Q$ is an increasing function as time goes by.
So, the dynamics with no interactions and no invasions in our model show one dominant species continues to grow larger and the others continue to grow smaller until $F$ reaches to~1 (see~Eq.\eqref{eq:cases}).

\subsection{With interactions ($\sigma > 0$)}
\label{subapp:w_interactions}

If we consider $\sigma$ to be nonzero, it is more difficult to understand.
Although the difficulty, we can compare the dynamics with a non-interacting system.
Then, Eq.\eqref{eq:dFdt_dqdt} shows simply as
    \begin{equation}
    \begin{split}
    & \frac{dF}{dt} = F^2 \left[
        \left(Q+\left\langle g_i(\textbf{q})\right\rangle_{\textbf{q}}\right)(1-F)
        +\left\langle D_i(\textbf{q})\right\rangle_{\textbf{q}}
        \right] \:, \\
    & \frac{dq_i}{dt} = Fq_i \left[ 
        \left(q_i+g_i(\textbf{q})-Q-\left\langle g_i(\textbf{q}) \right\rangle_{\textbf{q}}\right)(1-F)
        +D_i(\textbf{q})-\left\langle D_i(\textbf{q}) \right\rangle_{\textbf{q}}
        \right] \:.
    \end{split}
    \label{eq:dFdt_dqdt_1}
    \end{equation}
Due to stability, the stable fixed point $F^*$ is
    \begin{equation}
    F^* = 1+\frac{\left\langle D_i(\textbf{q}) \right\rangle_{\textbf{q}}}{Q+\left\langle g_i(\textbf{q}) \right\rangle_{\textbf{q}}}
    = 1-\delta(\sigma) \:,
    \label{eq:F_fp}
    \end{equation}
where $\delta(\sigma) \equiv - \frac{\left\langle D_i(\textbf{q}) \right\rangle_{\textbf{q}}}{Q+\left\langle g_i(\textbf{q}) \right\rangle_{\textbf{q}}}$ and $\delta(\sigma)$ is greater than or equal to zero because $w_{ij}^-$ is negative for all i and j. 
Then, to substitute $F^*$ as Eq.\eqref{eq:F_fp}, we can get that the fixed points are zero and
    \begin{equation}
    \begin{split}
    {q_i}^* & = \left(\frac{-D_i(\textbf{q}) + \left\langle D_i(\textbf{q}) \right\rangle_{\textbf{q}}}{1-F}\right)
        -g_i(\textbf{q})+Q+\left\langle g_i(\textbf{q}) \right\rangle_{\textbf{q}} \\
    & = Q + \left(\frac{-D_i(\textbf{q}) + \left\langle D_i(\textbf{q}) \right\rangle_{\textbf{q}}}{1-F}\right)
        -g_i(\textbf{q})+\left\langle g_i(\textbf{q}) \right\rangle_{\textbf{q}} \\
    & = Q + \left(-D_i(\textbf{q}) \right)\frac{1}{\delta(\sigma)}
        +\frac{\left\langle D_i(\textbf{q}) \right\rangle_{\textbf{q}}}{\delta(\sigma)}
        -g_i(\textbf{q})+\left\langle g_i(\textbf{q}) \right\rangle_{\textbf{q}} \\
    & = Q + \left(-D_i(\textbf{q}) \right)\frac{1}{\delta(\sigma)}
        -Q - \left\langle g_i(\textbf{q}) \right\rangle_{\textbf{q}}
        -g_i(\textbf{q})+\left\langle g_i(\textbf{q}) \right\rangle_{\textbf{q}} \\
    & = Q - Q +\left(-D_i(\textbf{q}) \right)\frac{1}{\delta(\sigma)}-g_i(\textbf{q}) \\
    & = Q - Q +\left(Q+\left\langle g_i(\textbf{q}) \right\rangle_{\textbf{q}} \right)\frac{D_i(\textbf{q})}{\left\langle D_i(\textbf{q}) \right\rangle_{\textbf{q}}} - g_i(\textbf{q}) \\
    & = Q + \left( \frac{D_i(\textbf{q})-\left\langle D_i(\textbf{q}) \right\rangle_{\textbf{q}}}
        {\left\langle D_i(\textbf{q}) \right\rangle_{\textbf{q}}} \right) Q 
        +\frac{\left\langle g_i(\textbf{q}) \right\rangle_{\textbf{q}} D_i(\textbf{q})
        -g_i(\textbf{q}) \left\langle D_i(\textbf{q}) \right\rangle_{\textbf{q}}}
        {\left\langle D_i(\textbf{q}) \right\rangle_{\textbf{q}}} \\
    & \approx Q + \left( \frac{D_i(\textbf{q})-\left\langle D_i(\textbf{q}) \right\rangle_{\textbf{q}}}
        {\left\langle D_i(\textbf{q}) \right\rangle_{\textbf{q}}} \right) Q \:,
    \end{split}
    \label{eq:q_fp}
    \end{equation}
where we ignore the higher order term of $\sigma$.
Therefore, nonzero fixed points are
    \begin{equation}
    \begin{split}
    \therefore\;\;
    & F^* = 1-\delta(\sigma) \:, \\
    & {q_i}^* \approx Q + \left( \frac{D_i(\textbf{q})-\left\langle D_i(\textbf{q}) \right\rangle_{\textbf{q}}}
        {\left\langle D_i(\textbf{q}) \right\rangle_{\textbf{q}}} \right) Q \:.
    \end{split}        
    \label{eq:Fq_fp}
    \end{equation}
In our assumption ($F<1$), the unstable fixed point ${q_i}^*$ depends on the interaction strength $\sigma$.
If the magnitude of $D_i(\textbf{q})$ is larger than the mean, the second term of the RHS of ${q_i}^*$ in Eq.\eqref{eq:Fq_fp} has a higher positive value.
That is, the unstable fixed point is a larger value than the others. 
So, the abundance of species $i$ is attracted to zero if the abundance is not enough to be large.
In the other case, the abundances continue to grow larger easily because the value of the unstable fixed point is small.
The difference between $D_i(\textbf{q})$ and its mean depends on the interaction strength $\sigma$ because it relates to the variance of $D_i(\textbf{q})$.
So, the bigger $\sigma$ is, the easier a dominant species appears. 

In our model, the invasion process plays the role of interrupting that $F$ dynamics reach equilibrium.
As a result, a dominant species can appear even though $\sigma$ is too small because $q_i$, $F$, and $Q$ cannot reach equilibrium.  

\newpage
\bibliography{bibliography}